# Microscopic kinetics pathway of salt crystallization in graphene nanocapillaries

Lifen Wang[1,2,*], Ji Chen[3], Stephen J. Cox[4,†], Lei Liu[5,‡], Gabriele C. Sosso[6], Ning Li[7,8], Peng Gao[3,7,8], Angelos Michaelides[4,9,10], Enge Wang[1,2,7,11], Xuedong Bai[1,2,12,§]

[1]Beijing National Laboratory for Condensed Matter Physics, Institute of Physics, Chinese Academy of Sciences, Beijing 100190, China.

[2]Songshan Lake Laboratory for Materials Science, Dongguan 523000, China.

[3]School of Physics and the Collaborative Innovation Center of Quantum Matters, Peking University, Beijing 100871, China.

[4]Yusuf Hamied Department of Chemistry, University of Cambridge, Lensfield Road, Cambridge CB2 1EW, United Kingdom.

[5]Department of Materials Science and Engineering, College of Engineering, Peking University, Beijing 100871, China.

[6]Department of Chemistry and Centre for Scientific Computing, University of Warwick, Coventry CV4 7AL, United Kingdom.

[7]International Center for Quantum Materials, School of Physics, Peking University, Beijing 100871, China.

[8]Electron Microscopy Laboratory, School of Physics, Peking University, Beijing 100871, China.

[9]Department of Physics and Astronomy, and Thomas Young Centre, University College London, London WC1E 6BT, United Kingdom.

[10]London Centre for Nanotechnology, University College London, London WC1H 0AH, United Kingdom.

[11]School of Physics, Liaoning University, Shenyang 110036, China.

[12]School of Physical Sciences, University of Chinese Academy of Sciences, Beijing 100190, China.

The fundamental understanding of crystallization, in terms of microscopic kinetic and thermodynamic details, remains a key challenge in the physical sciences. Here, by using *in situ* graphene liquid cell transmission electron microscopy, we reveal the atomistic mechanism of NaCl crystallization from solutions confined within graphene cells. We find that rock salt NaCl forms with a peculiar hexagonal morphology. We also see the emergence of a transitory graphite-like phase, which may act as an intermediate in a two-step pathway. With the aid of density functional theory calculations, we propose that these observations result from a delicate balance between the substrate-solute interaction and thermodynamics under confinement. Our results highlight the impact of confinement on both the kinetics and thermodynamics of crystallization, offering new insights into heterogeneous crystallization theory and a potential avenue for materials design.

Understanding and controlling the crystallization of materials from solution is of essential importance in various scientific and technological disciplines, including materials science, biology, geology, and atmospheric science [1,2]. In nature, complex phenomena such as cloud precipitation, biomineralization, and rock formation are associated with crystallization from solution [3], while industrially, solution-based methods offer a relatively simple and low-cost option for mass production [4]. This has motivated a large number of studies aimed at controlling the dynamics of nucleation, for example, the nucleation density, growth rate, and properties of crystals. For instance, by using additives, metals with finer grains, strengthened mechanical properties, and greater resistance to e.g. salt damage, have been obtained in metallurgy [5,6].

Classical nucleation theory (CNT) gives a largely reasonable description of nucleation and crystallization. However, important questions remain, e.g., whether or not the stable phase nucleates from solution in a single- vs. multi-step fashion involving intermediate phases [7-12]. While still challenging, *in situ* graphene liquid cell (GLC) imaging techniques provide a means to elucidate much needed microscopic insights into crystallization mechanisms [13]. Based on the assumption that graphene only interacts weakly with solution, the impact of the GLC on crystallization is often simply interpreted by effects due to reduced dimensionality and nanocapillary pressure, e.g. in studies of confined water [14-16]. However, a delicate balance between the substrate-solute interaction [17-19], the solute-solvent interaction, and thermodynamics under confinement [20] offers a new degree of freedom to modulate the crystallization pathway.

Conventional understanding suggests that NaCl follows a one-step classical nucleation pathway and grows into its conventional cubic rock salt structure (B1-NaCl) [21,22]. Using atomic-resolution *in situ* transmission electron microscopy (TEM), we reveal that, in a GLC, NaCl unexpectedly crystallizes into hexagonal-shaped crystallites, which predominantly expose their {110} facets instead of the conventional {100} facets. More surprisingly, a graphitic-like hexagonal NaCl phase (h-NaCl) [23] appears as an intermediate structure in the crystallization process, hinting at a non-classical nucleation pathway of NaCl in the GLC. Combined with density functional theory (DFT) calculations and control experiments, we highlight the importance of the interaction between the nascent crystallites and the graphene substrate, which could be considered as a kinetic approach to stabilize the hidden metastable phase and even as means to effect non-classical nucleation under confinement more generally.

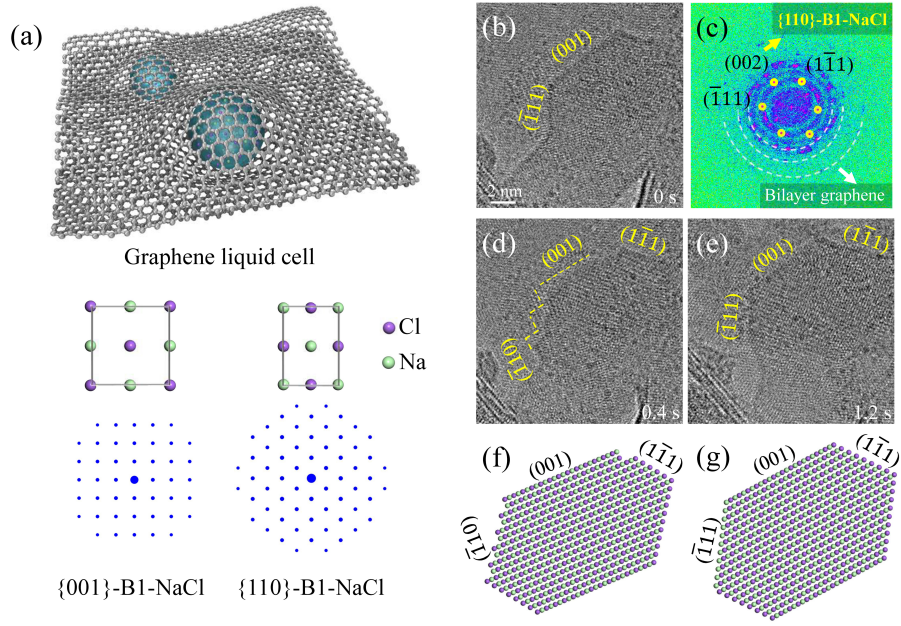

FIG. 1. Growth of the hexagonal-shaped B1-NaCl island. (a) Schematic of droplets encapsulated in a GLC, and the B1-NaCl crystal lattice viewed along the [001] and [110] crystallographic directions along with their corresponding diffraction patterns. (b) TEM image of a NaCl crystal grown in the GLC, showing low-index facets with 120° angles. (c) Corresponding diffractogram of the TEM image in (b), indicating NaCl oriented along [110]. (d), (e) Snapshots of the NaCl crystal growth process. While the (001) plane grows continuously, the ($\bar{1}$11) surface shows a saw-toothed plane composed of (001) and ($\bar{1}$10) facets, indicating that the layer-by-layer lateral growth of (001) and ($\bar{1}$10) facets dominate. (f), (g) Corresponding schematics of the TEM images shown in (d), (e).

Our experimental setup, comprising a quasi-two-dimensional graphene nanocell, is illustrated in Fig. 1(a). Given the higher electron scattering power of the saturated NaCl solution, we can identify the solution-encapsulated cells in the suspended TEM grid holes (see Fig. S1 in the Supplementary Material [24]). Figures 1(b), (d) and (e) show sequential high-resolution TEM images from one crystallization event (Video 1). The corresponding diffractogram is shown in Fig. 1(c), which demonstrates that the nanocrystal has a B1-NaCl structure along the {110}-zone axis (referred to as '{110}-B1-NaCl' hereafter). On-site electron energy loss spectroscopy (EELS) further confirms the NaCl composition (Fig. S2). The imaged nanocrystals show clear {100} and {111} facets with 120° angles, while {110} facets also evolve as transient side facets during the growth of this hexagonally shaped nanocrystal; as NaCl usually crystallizes into a cubic morphology, this observation comes as somewhat of a surprise. Moreover, this observation appears to be a feature of the GLC environment, as control experiments using open $SiN_x$ cells produced B1 crystals with their usual cubic morphology (Fig. S3).

Although the stable B1 phase ultimately forms in our experiments, a different structural phase of NaCl is occasionally seen to form during the crystallization process. Figures 2 (a)-(d) show high-resolution TEM images along with the corresponding diffractograms. In the diffractograms, the graphene sheets with a rotation angle of 30° give rise to spots marked by outer white dashed circles (Fig. S4). In addition to these, six spots all with the same vector length from the center can be clearly recognized (yellow circles). These additional spots originate from the NaCl crystals, which is confirmed using EELS (Fig. 2(e)), and distribute evenly on the circle. Importantly, they differ from the first-order diffraction pattern of a '{111}-B1-NaCl' crystal; the {111} planes of B1-NaCl are hexagonally close-packed so that {111}-B1-NaCl also has six equivalent first-order diffraction spots, but the vector length would be much larger. The six-fold symmetry of spots in reciprocal space thus indicates the formation of a hexagonal crystal structure of NaCl (h-NaCl), rather than the well-known B1 phase. Moreover, we carefully analyzed the relative angles between the graphene and the NaCl lattice and find the NaCl is not in registry with the graphene substrate.

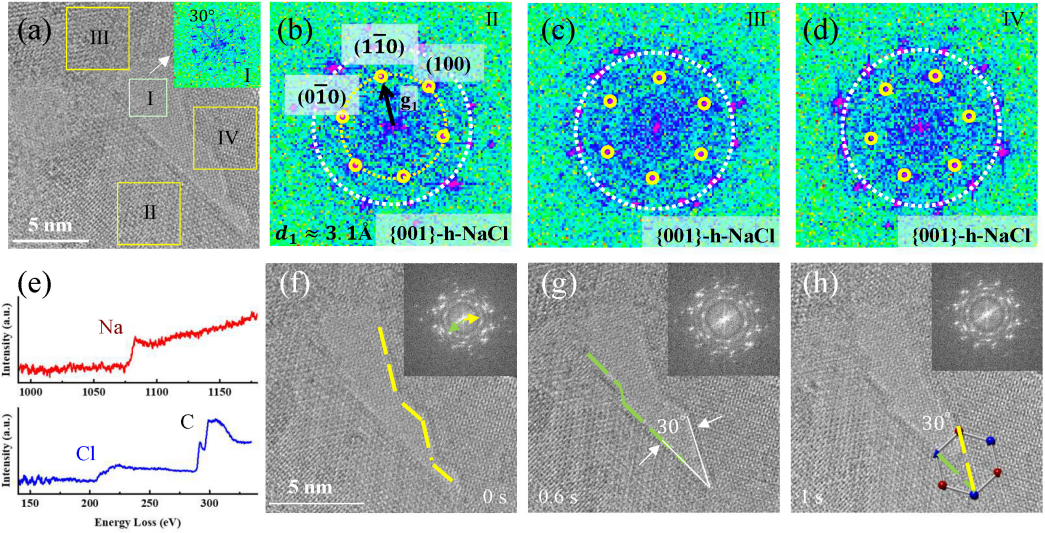

FIG. 2. Formation of the h-NaCl phase. (a) TEM image of NaCl crystals grown in a GLC, showing an in-plane hexagonal lattice. Inset: diffractogram of a graphene-only area with a rotation angle of approx. 30°. (b)-(d) The corresponding diffractograms of selected regions (yellow boxes) in (a), clearly showing six-fold symmetry (inner yellow circles). The rotation angles between NaCl and graphene (outer white dashed circle) varies. (e) EELS of the pocket area, showing signals only from C, Cl, and Na. (f)-(h) Sequential TEM images of h-NaCl crystal growth. The two predominant surfaces have an angle of 30°.

Our assignment of this transient structure as h-NaCl is corroborated by an analysis of its facet growth dynamics. Figs. 2(f)-(h) show a sequence of TEM images from the longest-lived h-NaCl crystal. This allows us to track the growth of its facets (Videos 2, 3). We find two predominant surfaces oriented 30° with respect to each other, as highlighted by the dashed lines. This is clearly distinct from the 120° angle seen for {110}-B1-NaCl, and the 90° angle expected for a typical B1 crystal exposing its {100} facets. The facet growth shown in Fig. 2 resembles other hexagonal materials such as graphene and hexagonal ice, where competition between the so-called zigzag and armchair edges is likely to evolve. Both the diffractograms and the facet growth dynamics are consistent with characterizing this transient structure as h-NaCl.

The observation of h-NaCl formation is intriguing and not expected based on conventional understanding of NaCl. To ascertain what role, if any, h-NaCl plays in the crystallization process, in Fig. 3 we present a sequence of TEM images depicting an entire crystallization event. Insets show the corresponding diffractograms. Initially, only a dark region corresponding to the encapsulated liquid is observed, and no sign of crystallization was seen (Fig. 3(a)). In the early stages of crystallization, h-NaCl forms with a well-defined six-fold symmetry (Fig. 3(b)), which after approx. 3s, begins to shrink (Fig. 3(c)), along with the emergence of B1 crystallites (Fig. 3(d)). These B1 crystallites subsequently dominate the crystallization process, leading to the formation of a {110}-B1-NaCl nanocrystal showing well-marked facets with 120° angles (Fig. 3(e)). From these images, we cannot determine if the B1 crystallites have formed via a solid-to-solid transition, or if they have formed independently of h-NaCl. Thus, it remains an open question whether h-NaCl acts simply as an early—but ultimately unsuccessful—competitor of B1-NaCl (similar to recent observations in protein crystallization [41]), or if it acts as an intermediatory phase in a two-step mechanism [42]. The crystallization event shown in Fig. 3 typifies three out of the five events seen in our experiments (Videos 4-7). In the remaining two events, {110}-B1-NaCl is seen to form without the detection of h-NaCl, although we cannot preclude e.g. the prior formation of a thin layer of h-NaCl. On balance, our results lean toward h-NaCl acting as an unsuccessful competitor to direct B1-NaCl formation, though we cannot definitively rule out a two-step mechanism. In any case, it appears that the crystallization pathway is altered in a GLC environment, even qualitatively, from that in bulk solution.

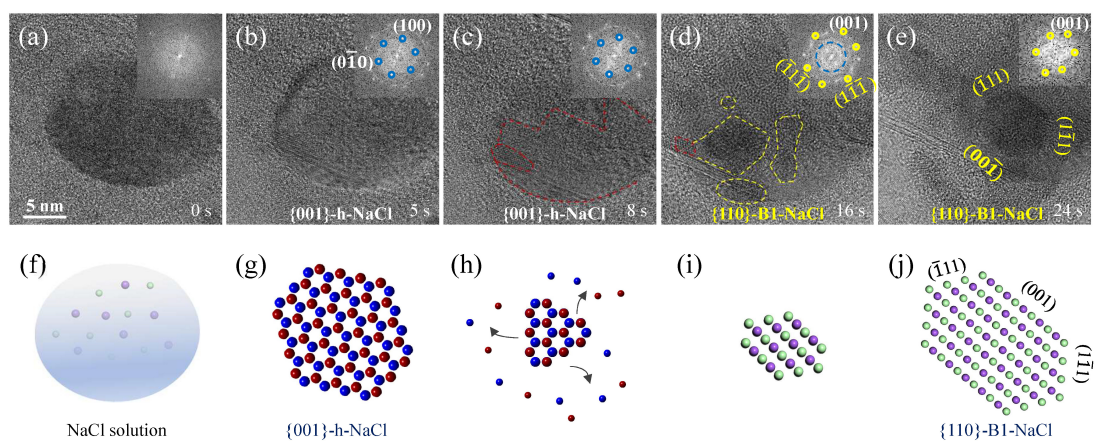

Fig. 3. Transformation from h-NaCl to {110}-B1-NaCl. (a)-(e) Sequential TEM images of a graphene pocket. Dark contrast in the middle of the image indicates the area of the trapped solution. Initially there is no crystal signal from the pocket ((a), inset). After 5s nuclei with the hexagonal structure fill the whole graphene pocket with a uniform lattice (b). Inset of (b) shows spots with six-fold symmetry. In the next stage the hexagonal structure shrinks (c), followed by transformation to B1-NaCl nuclei (d). Dashed red and yellow lines highlight the h-NaCl and B1-NaCl regions, respectively. Eventually, a large {110}-B1-NaCl crystal with a hexagonal shape is observed (e). (f)-(j) Schematics of the corresponding processes in (a)-(e).

Previous theoretical studies predict that h-NaCl is more stable than the B1 phase at large negative pressures, when it is a few layers thin, or when supported by a substrate [43-45]. Previous experimental studies have estimated high pressures (GPa) in GLCs. The fact that B1-NaCl (albeit with an exotic morphology) ultimately forms at the expense of h-NaCl suggests the former is stable while the latter is metastable. It therefore seems unlikely that pressure underlies the formation of h-NaCl. In order to further understand effects due to solvation and the interactions between the NaCl crystal and graphene, we have performed DFT calculations for three types of cluster, shown schematically in Figs. 4(a)-(c): '{001}-B1' clusters, '{110}-B1' clusters, and 'h-clusters'. In Fig. 4(d) we present calculated formation energies $e_f$ per formula unit for different sized clusters. These calculations have been performed both in vacuum and with an implicit solvent model [46]. We see that the {001}-B1 clusters are significantly more stable than the {110}-B1 clusters in vacuum. In solution however, the calculations with the implicit solvent model [25,30] suggest that the {110}-B1 clusters are marginally more stable, which is also reflected in calculations with extended surfaces (Fig. S6). As these results have been obtained with an implicit solvent model we exercise caution, and simply take this as indicative that the difference in surface energies of {001} and {110} facets is greatly reduced in solution compared to their stark energy difference in vacuum. More importantly, we find that the {110}-B1 clusters interact much more favorably with graphene than do the {001}-B1 clusters, as shown in Fig. 4(e). This suggests that the formation of {110}-B1-NaCl nanocrystals (Fig. 1) may be driven by a combination solvation effects and a favorable interaction between {110} facets and graphene.

In Fig. 4(d) we also see that, while $e_f$ for the bulk hexagonal crystal (dotted line) is far higher compared to that of the B1 structure (dashed line), the h-NaCl clusters are energetically similar to the B1 clusters. This is the case both in vacuum and with solvent. For very small cluster sizes of h-NaCl, its interaction with graphene (Fig. 4(e)) is comparable to the {110}-B1 clusters, though it becomes relatively less strong as size increases. During the initial stages of crystallization, we suggest the formation of the h-NaCl crystal is not disfavored on energetic grounds. As the crystal becomes larger, the preference for the B1 structure increases. These calculations are consistent with our experimental observation that the {110}-B1-NaCl nanocrystal ultimately forms.

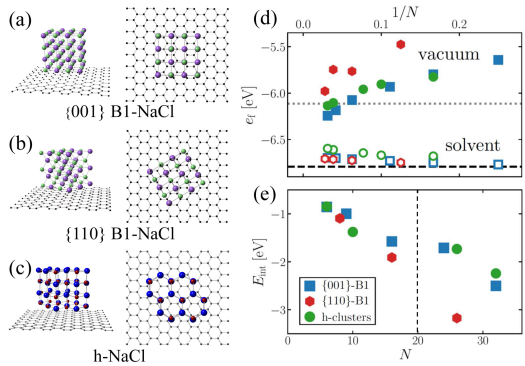

FIG. 4. Energetics of NaCl clusters. (a-c) Schematic representations of {001} B1, {110}-B1 and h-clusters bound to graphene, as indicated at the bottom of each panel. (d) Formation energy per formula unit ($e_f$) *vs.* $1/N$. In vacuum, {001}-B1 clusters (squares) are more stable than {110}-B1 clusters (hexagons) for finite N. In solvent, the two structures are energetically similar. While the bulk energy of the hexagonal structure (dotted line) is significantly higher than that of the B1 structure (dashed line), the energy difference is far less pronounced for h-clusters (circles). (e) Interaction energies with graphene *vs.* N. The {110}-B1 clusters interact much more favorably with graphene than do the {001}-B1 clusters. The vertical dashed line separates small clusters (all clusters interact with graphene similarly) and large clusters ({110}-B1 cluster interacts more strongly with graphene).

There is increasing evidence that nucleation occurs non-classically; notable experimentally observed examples include proteins [7], minerals [47], colloids [48] and polymeric solutions [1]. NaCl, a salt with a simple structure, is believed to follow CNT and shows cubic morphology. Our results suggest elements of non-classical nucleation extend to NaCl, with the possibility of crystallization via an intermediate metastable crystalline phase: this appears to be distinct from the non-classical mechanism reported at very high supersaturations [49]. More importantly, our results show that the crystallization pathway in solution can be engineered by the interaction between the crystallites and substrates, i.e. the unexpected transient formation of h-NaCl as a kinetic stable state in the nano-sized capillaries, compared with the conventional cubic structure in the micron-sized open cell. In principle, this approach to achieve a non-classical nucleation pathway could be readily extended to other systems [1,2,12].

Furthermore, in addition to the revealed crystallization mechanism, our findings clearly show that confinement in a GLC alters crystallization of NaCl both in terms of morphology and intermediate/transitory metastable phases without involving ultrahigh pressure. This opens up exciting possibilities in nanocrystal design; for example, the morphology controlling ability of the GLC could be useful for catalysis, where the catalytic behavior of a material sensitively depends on exposed facets [50]. Moreover, there has been growing interest in metastable crystal structures of functional materials, including the III-V compound semiconductors and transition metal dichalcogenides [51,52]. Future studies may therefore use graphene-confined cells to grow these materials and to discover unknown metastable phases.

We acknowledge G. Gu from University of Tennessee for constructive and fruitful discussions. Funding: This work was supported by the Program from Chinese Academy of Sciences (Y8K5261B11, ZDYZ2015-1, XDB30000000, XDB33030200 and XDB07030100), National Natural Science Foundation (11974388, 11974024, 11974001, U1932153, 21872172, 21773303, 51991340 and 51991344), National Key R&D Program under Grant No. 2019YFA0307801, and Beijing Natural Science Foundation (2192022, Z190011). S.J.C. is supported by a Royal Commission for the Exhibition of 1851 Research Fellowship. L.W. is grateful for the support from the Youth Innovation Promotion Association of CAS (2020009). We thank the TianHe-1A supercomputer, the High Performance Computing Platform of Peking University, and the Platform for Data Driven Computational Materials Discovery of the Songshan Lake Materials Lab for computational resources. We are grateful to the UK Materials and Molecular Modelling Hub for computational resources, which is partially funded by EPSRC (EP/P020194/1 and EP/T022213/1).

L. W., J. C., and S. J. C. contributed equally to this work.
*wanglf@iphy.ac.cn;
†sjc236@cam.ac.uk

‡l_liu@pku.edu.cn

§xdbai@iphy.ac.cn

# Supplemental Material for

# Microscopic kinetics pathway of salt crystallization in graphene nanocapillaries

Lifen Wang[1,2,*], Ji Chen[3], Stephen J. Cox[4,†], Lei Liu[5,‡], Gabriele C. Sosso[6], Ning Li[7,8], Peng Gao[3,7,8], Angelos Michaelides[4,9,10], Enge Wang[1,2,7,11], Xuedong Bai[1,2,12,§]

[1]Beijing National Laboratory for Condensed Matter Physics, Institute of Physics, Chinese Academy of Sciences, Beijing 100190, China.

[2]Songshan Lake Laboratory for Materials Science, Dongguan 523000, China.

[3]School of Physics and the Collaborative Innovation Center of Quantum Matters, Peking University, Beijing 100871, China.

[4]Yusuf Hamied Department of Chemistry, University of Cambridge, Lensfield Road, Cambridge CB2 1EW, United Kingdom.

[5]Department of Materials Science and Engineering, College of Engineering, Peking University, Beijing 100871, China.

[6]Department of Chemistry and Centre for Scientific Computing, University of Warwick, Coventry CV4 7AL, United Kingdom.

[7]International Center for Quantum Materials, School of Physics, Peking University, Beijing 100871, China.

[8]Electron Microscopy Laboratory, School of Physics, Peking University, Beijing 100871, China.

[9]Department of Physics and Astronomy, and Thomas Young Centre, University College London, London WC1E 6BT, United Kingdom.

[10]London Centre for Nanotechnology, University College London, London WC1H 0AH, United Kingdom.

[11]School of Physics, Liaoning University, Shenyang 110036, China

[12]School of Physical Sciences, University of Chinese Academy of Sciences, Beijing 100190, China.

**This PDF file includes:**

  Figures S1 to S7

  References

**Other Supporting Materials for this manuscript include the following:**

  Videos 1-9

**Methods**

**Material preparation and *in situ* TEM imaging**

Monolayer graphene was grown on Cu foils by chemical vapor deposition. Graphene liquid cells are fabricated by transferring monolayer graphene onto a Quantifoil TEM grid, exposing it to one micro-liter of saturated NaCl solution, covering it with another graphene loaded TEM grid, and leaving the cell to dry in a pump-station overnight. Transparent overlapping holes in the carbon membranes of the gold grid allow electron illumination of graphene cells thus formed (Fig. S1). The TEM imaging and EELS were performed using the aberration corrected JEOL JEM-ARM300F microscope operated at 80 KV. The electron beam dose rate is minimized to $10^3 e^-/Å^2 s$ to ensure the imaging life time of the graphene cell. Typically for the *in situ* TEM imaging, the spherical aberration was corrected to ~-5μm. The 2-fold astigmatism was corrected to ~0 nm. The coma and 3-fold astigmatism were corrected to several nanometers.

**Control experiment with water**

To evaluate the intrinsic capabilities of the graphene liquid cell, we first perform a control experiment in which purified water is encapsulated. As shown in Fig. S1c, the two suspended graphene sheets have a rotation angle of ~15.2° and the lateral size of the overlapping region is several hundreds of nanometers. Droplets encapsulated in nanoscale graphene pockets feature with a dark contrast in the resulting TEM image. Fig. S3a shows the typical overview image of the graphene liquid cells with deionized water encapsulated. Nanometer-sized pockets form with features typical of a fluid, indicating good adhesion between two monolayer graphene sheets (Supporting Video 8). Since the graphene cell is assembled in an ambient environment, air contamination is inevitable, thus the growth of bubbles on the hydrophobic graphene surface is observed. The gas bubbles diffuse out in seconds along the interface, owing to graphene's gas impermeability. The measured in-plane diameter size of the isolated pockets ranges from 2 nm to 26 nm in the control experiments. No notable crystalline features other than the graphene lattice are observed.

**Control experiment with saturated NaCl solution**

In comparison with the crystallization behavior of NaCl in graphene liquid cell, a one-inlet flow stage for *in situ* liquid cell TEM and silicon cell chips (Hummingbird Scientific) were used for observations of saturated NaCl solution. Saturated NaCl solutions were pumped through and around the cell. The fluidic solution is sealed from the high vacuum in the TEM column by two Viton O-rings. The liquid cells were assembled by overlapping suspended $SiN_x$ windows on bottom and top chips. As shown in Fig. S3 b-e and Supporting Video 9, NaCl crystallizes with the typical cubic morphology with the B1 structure in the open cell. Because of the relative open environment, the solution as well the cubic NaCl were expelled away by the electron beam radiolysis.

**DFT calculations**

Density functional theory calculations were performed using VASP [1,2]. In all calculations Gamma point was used for K-mesh sampling and the supercells were larger than 20 Å. The PAW pseudopotentials [3] were used with an energy cut-off 500 eV. The van der Waals inclusive optB86b-vdw exchange correlation functional [4,5] was used for all DFT calculations. Implicit solvation was applied with the VASPsol package [6,7], in which the relative permittivity of water was set to 80. In the calculations with implicit solvation, a 9-valence electron PAW potential was used for Na and the energy cut-off was increased to 800 eV.

The formation energy was defined as

$$e_f = (E_{cluster} - E_{Na} - E_{Cl})/N_{pair}$$

Where $E_{cluster}$ is the total energy of the cluster in vacuum and in implicit solvation, $E_{Na}$ and $E_{Cl}$ is the total energy of the Na and the Cl atom, respectively, and $N_{pair}$ is the number of ion pairs in the cluster.

The interaction of NaCl clusters with graphene was calculated as

$$E_{int} = E_{Gr+Cluster} - E_{Gr} - E_{Cluster}$$

Where $E_{Gr+Cluster}$ is the total energy of the adsorbed system, $E_{Gr}$ is the total energy of graphene and $E_{Cluster}$ is the total energy of a fixed cluster. To obtain $E_{Gr+Cluster}$, we also held the graphene fixed, and the separation between was varied until the lowest energy was identified. Curved graphene substrates were also considered (Fig. S7).

The surface energy (Fig. S6a) was calculated as

$$E_{surf} = \frac{E_{slab} - N_{pair} e_{bulk}}{2A},$$

where $E_{slab}$ is the total energy of the slab in contact with vacuum, $N_{pair}$ is the number of ion pairs in the slab, $e_{bulk}$ is the energy per ion pair of the bulk rock salt structure, and $A$ is the surface area.

**Molecular Dynamic simulations.**

Classical molecular dynamics simulations of NaCl in contact with water were performed using the simple point charge Joung-Cheatham [8] and SPC/E force fields [9]. The LAMMPS simulation package [2] was used. To first establish the surface energies of (001) and (110) surfaces of NaCl in vacuum, a bulk rock salt structure comprising 256 ion pairs was optimized at 1 atm using conjugate gradient with quadratic linesearch. The resulting simulation cell had a cubic cell dimension of 22.8589 Å. The crystal was then cleaved along the (001) surface, and the simulation cell was extended along the direction perpendicular to the surface (the *z* direction) was 91.4356 Å. This resulted in two (001) surfaces exposed to vacuum. The geometry of this slab was then optimized with the simulation cell fixed. A similar procedure was performed for the (110) surface.

We have found $E_{surf,001} = 1.398$ eV/nm$^2$ and $E_{surf,110} = 2.981$ eV/nm$^2$, for the (001) and (110) surfaces, respectively. While we do not expect the simple point charge model to be in quantitative agreement with the more sophisticated DFT approach, it correctly predicts that $E_{surf,110}$ is significantly greater than $E_{surf,001}$.

To estimate the effects of the solution environment, the vacuum regions were respectively filled with 1160 and 1830 water molecules for the (001) and (110) surfaces, and after suitable equilibration, molecular dynamics (MD) were performed for 1 ns at 298 K and 1 atm. Temperature was maintained with a Nosè-Hoover thermostat, and pressure was maintained with a Parrinello-Rahman barostat as implemented in LAMMPS [10-13]. (The cell dimensions were allowed to fluctuate independently.) Such a simulation was also performed for the bulk rock salt structure described above, along with a simulation of bulk water comprising

1024 molecules. In the case of the latter, an isotropic barostat was used such that a cubic simulation cell was maintained. Long ranged electrostatics were computed using particle-particle particle-mesh Ewald [14], with parameters chosen such that the root mean square error in the forces were a factor $10^5$ ($10^7$ for the vacuum surface energy calculations) smaller than the force between two unit charges separated by a distance of 1.0 Å [15]. Real space interactions were truncated and shifted at 10 Å. Dynamics were propagated using the velocity-Verlet algorithm with a 2 fs time step. The geometry of the water molecules was maintained with the RATTLE algorithm [16].

The surface enthalpy was then calculated as

$$H_{\text{surf}} = \frac{H_{\text{slab}} - N_{\text{pair}} h_{\text{NaCl}} - N_{\text{wat}} h_{\text{wat}}}{2\langle A \rangle},$$

where $H_{\text{slab}}$ is the total enthalpy of the slab in contact with water, $h_{\text{NaCl}}$ is the enthalpy per ion pair of bulk rock salt NaCl, $h_{\text{wat}}$ is the enthalpy per water molecule of bulk water, $N_{\text{wat}}$ is the number of water molecules in slab/water simulation, and $\langle A \rangle$ is the average surface area in contact with water. We have found $H_{\text{surf},001} = 0.679$ eV/nm$^2$ and $H_{\text{surf},110} = 0.788$ eV/nm$^2$ for the (001) and (110) surfaces, respectively. Clearly $H_{\text{surf}} \ll E_{\text{surf}}$ in both instances i.e. water stabilizes the surfaces compared to vacuum. Importantly, the extent of stabilization is far greater for the (110) surface: $H_{\text{surf},110} - H_{\text{surf},001} = 0.109$ eV/nm$^2$ vs. $E_{\text{surf},110} - E_{\text{surf},001} = 1.583$ eV/nm$^2$.

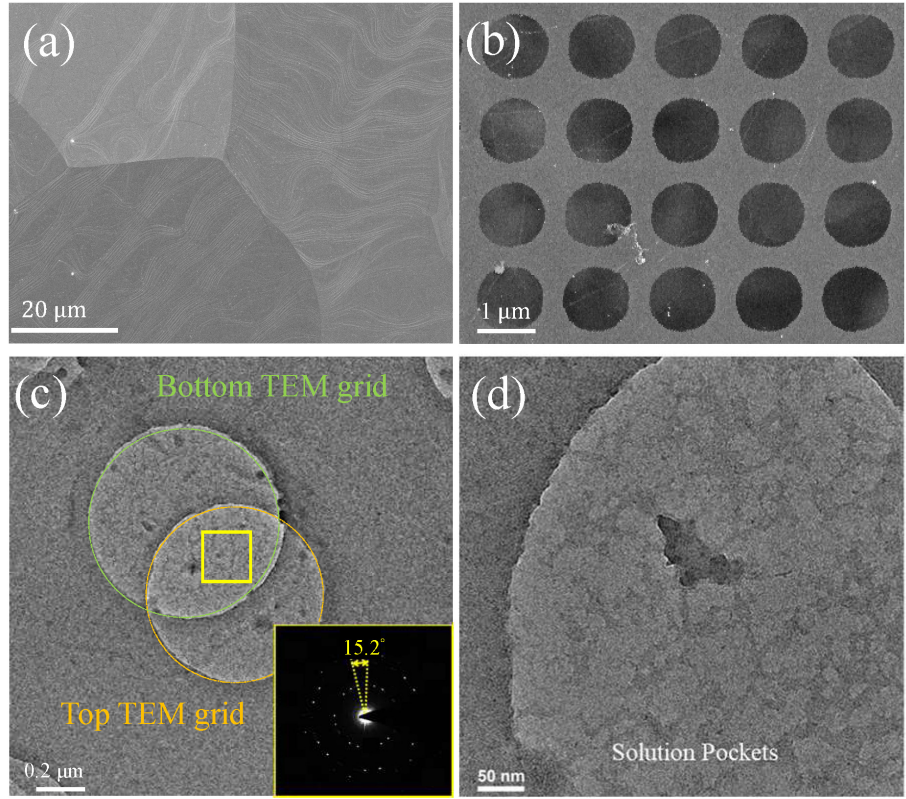

Fig. S1. Electron microscopy images of graphene liquid cells. (a) Scanning electron microscopy image of CVD grown monolayer graphene on copper substrate. (b) Scanning electron microscopy image of monolayer graphene covered TEM grid. (c) TEM image of graphene liquid cell fabricated by overlapping graphene-covered holes in membranes of two TEM grids. Inset: selected area diffraction pattern of the graphene cell indicating the 15.2° twisted bilayer graphene. (d) High-magnification TEM image of the graphene cell area in (c). Dark contrast highlight the locations of the cells with solution encapsulated.

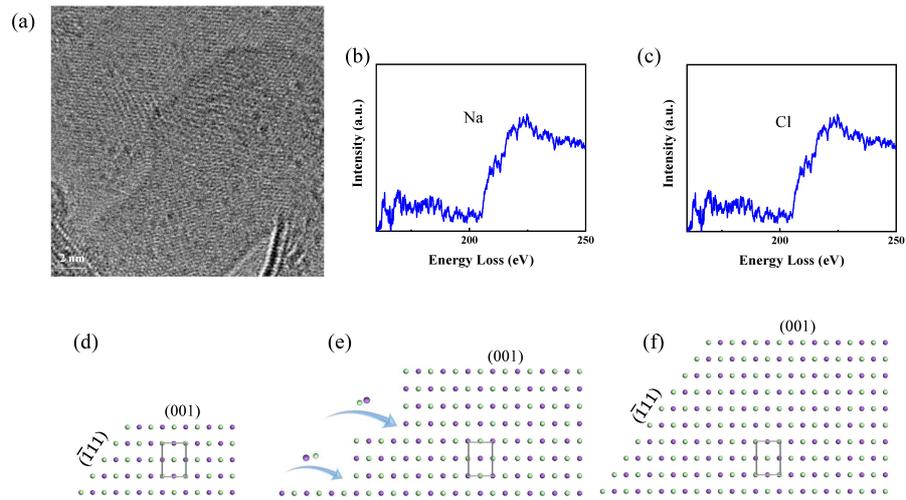

Fig. S2. Electron energy loss spectroscopy and growth steps illustration of the NaCl crystallite in a graphene cell. (a) Bright field TEM image of the crystallite formed in a graphene liquid cell. (b-c) Survey electron energy loss spectroscopy of the image area shows the composition of the crystallite is NaCl. (d-f) Schematic of the lateral growth of steps for the ($\bar{1}11$) surface of {110}-B1-NaCl.

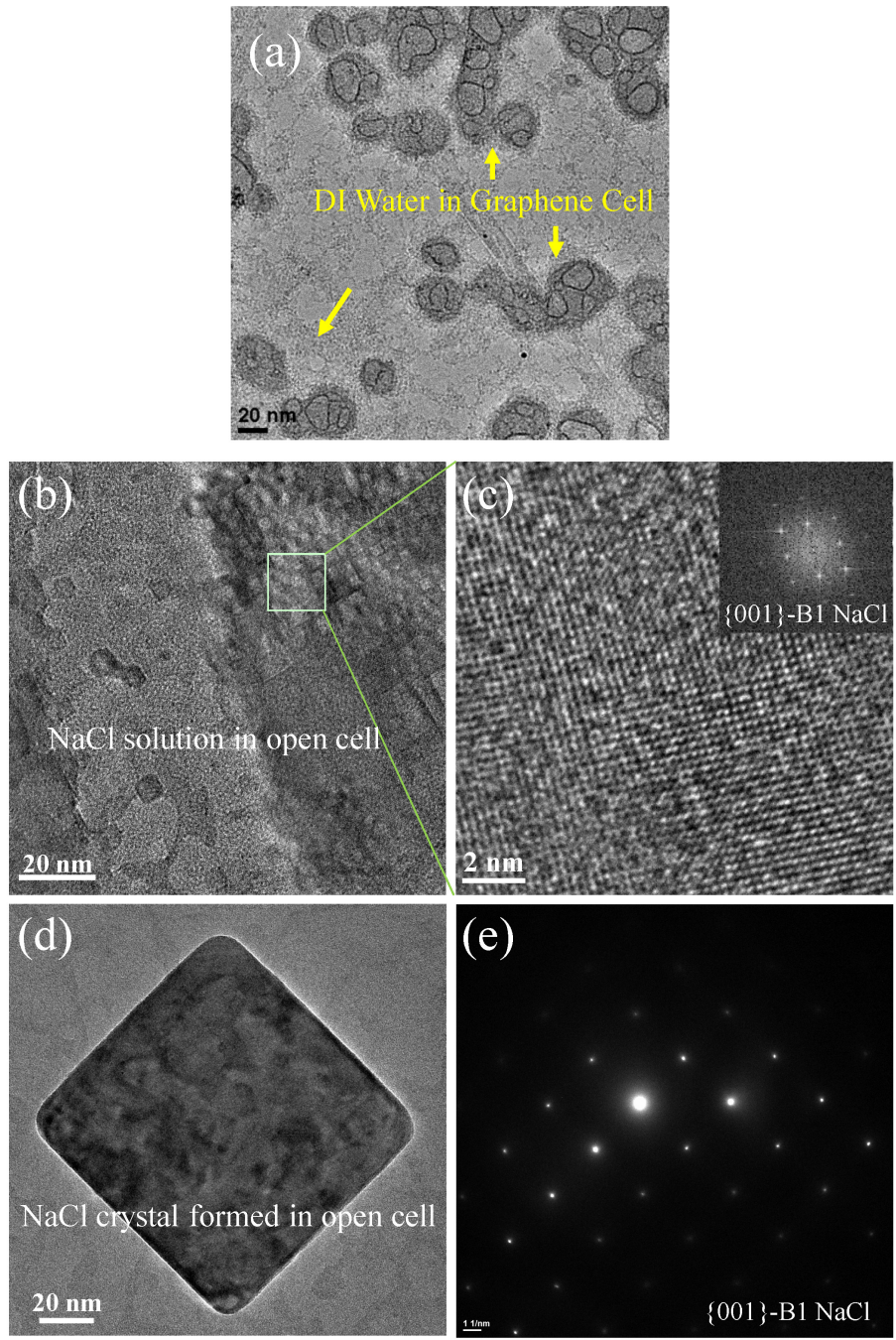

Fig. S3. TEM images of contrast cells. (a) TEM image of graphene liquid cell with deionized water encapsulated. The dark contrast pointed by yellow arrows show the water cells area. (b) TEM image of $SiN_x$ cell in which fluidic saturated NaCl solution is pumped in-between two 50 nm thick $SiN_x$ windows. (c) High-magnification TEM image of the cubic NaCl in (b) showing the B1 structure. (d, e) TEM image of a NaCl crystal formed from the saturated solution in the $SiN_x$ windows assembled liquid cell and the corresponding selected area diffraction pattern without stage tilting for the perfect zone-axis show the typical cubic morphology with B1 structure.

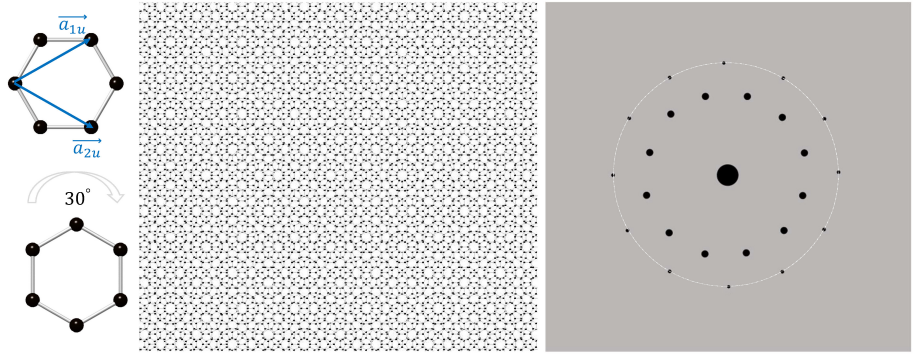

Fig. S4. Structure models of 30° twisted bilayer graphene and the corresponding pseudodiffraction pattern showing the featured quasicrystalline 12-fold symmetry. The white dashed circle marks the diffraction pattern from the two layers of graphene with a rotation angle of 30°. The inner 12-fold symmetry pseudodiffraction pattern comes from the Moiré pattern of the 30° twisted bilayer graphene.

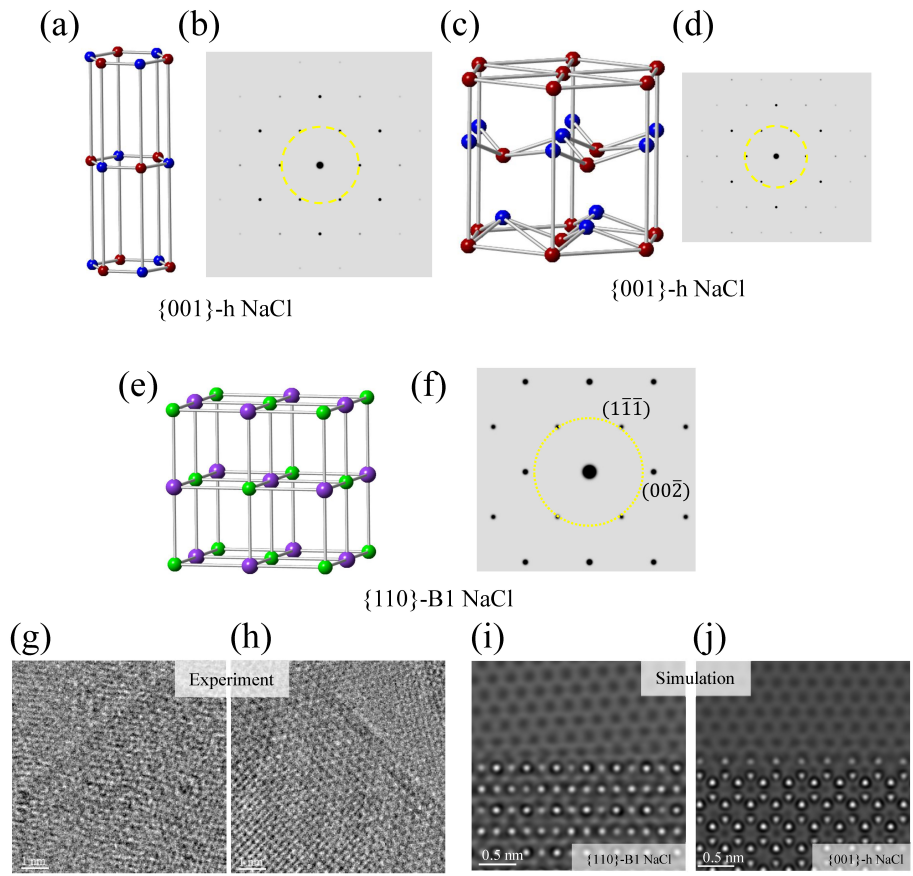

Fig. S5. Structure models of polymorph NaCl. (a) Structure model of hexagonal NaCl unit cell. (b) Single crystal diffraction pattern of [001] oriented hexagonal NaCl showing the six-fold symmetry. (c) Structure model of wurtzite NaCl unit cell. (d) Single crystal diffraction pattern of [001] oriented wurtzite NaCl showing the six-fold symmetry. (e) Structural model of face centered cubic NaCl. (f) Single crystal diffraction pattern of [110]-zone axis oriented face centered cubic NaCl. (g) TEM image of {110} B1-NaCl in a graphene liquid cell. (h) TEM image of h-NaCl in a graphene liquid cell. (i) Simulated TEM image for five-layers of {110}-B1-NaCl in a 4° twisted graphene cell. (j) Simulated TEM image for three layers of h-NaCl in a 30° twisted graphene cell.

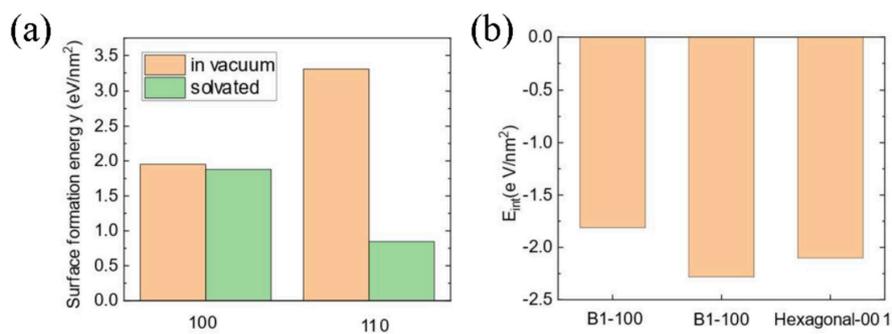

Fig. S6. DFT calculations of the surface formation energy and surface graphene interaction. (a) The surface formation energy of (100) and (110) surfaces in vacuum and in implicit solvent. The formation energy is defined as $E_f = (E_{slab} - N_{NaCl} \times E_{NaCl})/(2S)$, where $E_{slab}$ is the total energy of a periodic slab (in vacuum and in implicit solvation), $N_{NaCl}$ is the number of formula unit in the slab, $E_{NaCl}$ is the energy per formula unit of the bulk B1 NaCl, and $S$ is the surface area of one side of the slab. (b) The surface graphene interaction per area estimated considering big clusters adsorbing on graphene. $E_{int} = (E_{total} - E_{graphene} - E_{cluster})/S$, where $S$ is the area of the surface adsorbed on graphene. The clusters used contained 32, 36 and 48 NaCl for the B1-100, B1-110, and Hexagonal-001 clusters respectively.

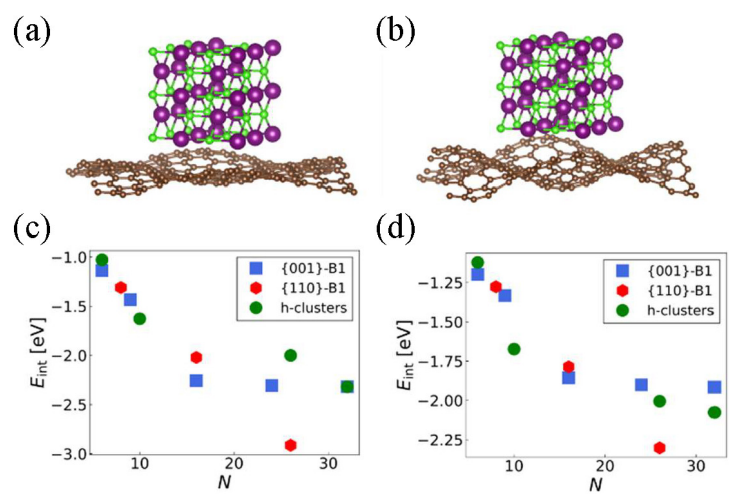

Fig. S7. DFT calculations of the curved surface graphene interaction. (a), (b) show structures of two curved (10% and 20%) graphene layers, each with an adsorbed NaCl cluster. The lateral dimension of the graphene layer is 21.3Å×19.7Å, hence the amplitude of the vertical distortion for the 10% and 20% curved layers is approximately 1.0 Å and 2.0 Å, respectively. (c) and (d) show the interaction energy between the curved graphene sheets (10% and 20%, respectively) and different NaCl clusters as a function of the size, namely the number of NaCl units.

**Supplementary Video 1**

Crystallization and growth of {001}-zone axis oriented fcc-NaCl in graphene cell. Images were collected at 5 frames per second (fps). Snapshots are shown in Fig. 1.

**Supplementary Video 2**

Crystallization and growth of hexagonal NaCl in graphene cell 1. Images were collected at 5 frames per second (fps).

**Supplementary Video 3**

Crystallization and growth of hexagonal NaCl in graphene cell 2. Images were collected at 5 frames per second (fps). Snapshots are shown in Fig. 2.

**Supplementary Video 4**

Low-magnification TEM imaging of the graphene cell with NaCl solutions encapsulated. Images were collected at 5 frames per second (fps).

**Supplementary Video 5**

High-magnification TEM imaging of the graphene cell in video 4 recording the formation of h-NaCl. Images were collected at 5 frames per second (fps). Snapshots are shown in Fig. 3.

**Supplementary Video 6**

*In situ* TEM imaging of the graphene cell following video 5, showing h-NaCl dissolve and subsequent {110}-B1-NaCl formation. Images were collected at 5 frames per second (fps). Snapshots are shown in Fig. 3.

**Supplementary Video 7**

Sequential TEM imaging of the graphene cell following video 6, showing the formation of {110}-B1-NaCl. Images were collected at 5 frames per second (fps). Snapshots are shown in Fig. 3.

**Supplementary Video 8**

Dynamic water and bubbles in graphene liquid cell under electron illumination. Images were collected at 5 frames per second (fps).

**Supplementary Video 9**

TEM imaging of the fluidic NaCl solution in $SiN_x$ liquid cell under electron illumination showing cubic crystallites. Images were collected at 5 frames per second (fps).